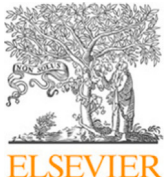
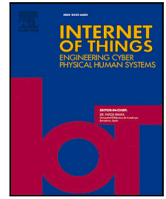

Contents lists available at ScienceDirect

# Internet of Things

journal homepage: www.elsevier.com/locate/iot

Review article

# An overview of IoT architectures, technologies, and existing open-source projects

Tomás Domínguez-Bolaño *, Omar Campos, Valentín Barral, Carlos J. Escudero, José A. García-Naya

*CITIC Research Center & Department of Computer Engineering, University of A Coruña, 15071, A Coruña, Spain*

A B S T R A C T

Today's needs for monitoring and control of different devices in organizations require an Internet of Things (IoT) platform that can integrate heterogeneous elements provided by multiple vendors and using different protocols, data formats and communication technologies. This article provides a comprehensive review of all the architectures, technologies, protocols and data formats most commonly used by existing IoT platforms. On this basis, a comparative analysis of the most widely used open source IoT platforms is presented. This exhaustive comparison is based on multiple characteristics that will be essential to select the platform that best suits the needs of each organization.

## 1. Introduction

Traditionally, companies and organizations used to have monitoring and action systems to cover different needs: heating, ventilation, and air conditioning (HVAC), leak detection (e.g., gases, liquids), presence, luminosity, etc. These systems have been implemented progressively, without coordination, causing dispersion between different departments: occupational risk prevention service, infrastructure service, information technology (IT), security, etc.

The explosion of the IoT has increased this dispersion, highlighting the need for platforms to coordinate and centralize these systems. Therefore, the concept of smart environments arises, where IoT platforms have been appearing applied to cities, university campuses, and industry. Although this concept already existed decades ago in industrial environments under the term supervisory control and data acquisition (SCADA) [1], this approach is not easily applicable to IoT because SCADA have always considered devices with very specific standards, a restricted range of manufacturers, and lacks mechanisms to ensure the security and confidentiality of information. This is not the case in the IoT world, where the number of manufacturers, standards, and scope of application are very large, hence requiring open architectures that do not fall into the error of the first proprietary systems in which both security and interoperability were completely missing.

In recent years, manufacturers of different devices have started to provide their own IoT solutions with a software as a service (SaaS) model, hence devices of a manufacturer will connect to its IoT service and to allow users to access the device's data or to interact with the devices remotely. However, these are usually proprietary and closed solutions, which aggravate the aforementioned dispersion and heterogeneity of monitoring systems. In addition, limitations would soon appear, for example:

- Maintenance of a new cloud platform with the typical problems exhibited by any new platform: access control management, potential software and hardware vulnerabilities, and periodic payment of licenses.

- Dependence on a new manufacturer. Any new devices would require hardware compatible with the cloud of the acquired solution, remaining captive to this solution without being able to integrate different types of devices.
- Limited access to sensor data for analysis. In a platform controlled by a single manufacturer the functionalities are predefined, and there is little to no control over them. It is desirable that the manufacturer provides an application programming interface (API) that allows external access to the data, but sometimes it does not exist, or the data provided does not follow a standard.

Thus, a general IoT platform that can be used for different devices without relying on a single manufacturer is desirable. There exist several proprietary enterprise IoT platforms on the market, usually provided under the SaaS model, that allow for connecting different devices, and collecting, storing, and analyzing their data. However, if the company or organization already has its own cloud and network infrastructure, another interesting possibility is to use some of the various free and open-source IoT platforms and projects available and deploy them directly on such infrastructure, avoiding the use of external cloud solutions.

Between different IoT platforms, the functionalities provided, their implementation and underlying technologies may differ significantly. Thus, when selecting an IoT platform we must consider the functional requirements, i.e., the actual requirements and functionalities that we need, and also the non-functional requirements, i.e., other aspects such as performance, cost, freedom of use, or development effort.

In the literature, several works have compared the existing IoT architectures, technologies, and platforms, or proposed comparison methodologies. In particular, in [2] a very extensive survey covering the topics of enabling technologies, protocols, cloud platforms, and applications was presented.

Regarding the IoT architectures and technologies, in [3] the authors compared the architectures of several IoT platforms and proposed a general IoT reference architecture. In [4], the authors presented a review of system architecture, software, communications, privacy, and security of IoT-based smart homes. Such a review includes comparisons of operating systems and communications technologies, but IoT platforms are not considered. In [5], another review of technologies is presented, which also includes a survey of proposed conceptual architectures (or models) and communication technologies, as well as challenges and open research issues. More recently, [6] provided an overview of the design and construction of IoT systems from the software engineering perspective, comparing the available hardware and software development tools for IoT systems.

With respect to the IoT platforms, in [7] the authors presented a survey of several IoT cloud platforms assessing application development, device management, system management, heterogeneity management, data management, tools for analysis, deployment, monitoring, visualization, and research applicability. A survey of several commercial cloud IoT platforms is available in [8] evaluating device management, integrations, security, protocols, analytics, and data visualizations of each IoT platform. In [9] the authors presented a study of several commercial cloud data platforms, assessing the platforms according to the functional requirements needed for the so-called Industry 4.0. The performance of both ThingsBoard and SiteWhere open-source IoT platforms is evaluated in [10]. A survey of different IoT platforms is presented in [11], in which the authors defined the following criteria for the comparison: topology (i.e., local-based, cloud-based, or hybrid), programming languages and application development, third-party support, extended protocol support, event handling, and security. A method for comparing IoT systems considering a probabilistic linguistic method is proposed in [12], whereas. [13] details a comparison framework and a methodology for IoT platforms based on functional and non-functional requirements, specifying in-depth taxonomies for these, and employing such a methodology to compare five commercial IoT platforms.

In this article we present a general overview of IoT architectures, technologies, and some of the most relevant open-source platforms and projects available as of today. The contributions of this paper can be summarized in the following points:

1. A comprehensive overview of key IoT architectures and technologies is presented in Section 2. First, in Section 2.1 we propose a detailed reference architecture of IoT systems based on the literature and previous IoT standards. Next, Sections 2.2 and 2.3 present an overview of data layer protocols and data formats used in IoT systems.
2. Existing IoT platforms are detailed in Section 3 considering two categories: commercial platforms as well as free and open-source platforms. The name and brief description of each of the most popular commercial platforms is included in Section 3.1. As mentioned above, although there are many works in the literature that describe and compare commercial IoT platforms in great detail, there are not as many works for the case of free and open-source platforms. To alleviate this shortcoming, Section 3.2 compares in detail several popular, free and open-source IoT platforms.
3. Section 4 provides an overview of other related free and open-source software projects that, although they may not be considered as IoT platforms, can provide important functionalities of the IoT platforms.
4. Finally, Section 5 summarizes the open-source platforms and software projects discussed in Sections 3 and 4.

## 2. IoT architectures and technologies

The IoT is built around IoT devices (the "things" in IoT), which are physical devices such as sensors and actuators that can exchange information and can be managed with some other external system, usually an IoT platform (i.e., a central server). An example of an IoT sensor is an air quality sensor connected to a central server where the user can see the current air quality reported by the sensor. An example of an IoT actuator is an air ventilation system connected to a central server that can be turned on or off by a user by just sending a message to the server. As seen in these examples, one of the core ideas of the IoT systems is the interaction between users and physical devices, i.e., on the one hand, sensors provide information to the user about the environment, and on the other hand, users can perform actions over other physical devices based on that information. Another key point of IoT systems is the possibility of developing "smart" applications for specific scenarios. These applications may involve different tasks such as data analysis, prediction, control, or monitoring and alerting users. For example, considering the previous examples, the user could configure rules in the IoT platform so that the ventilation system is automatically turned on or off depending on the values reported by the air quality sensor.

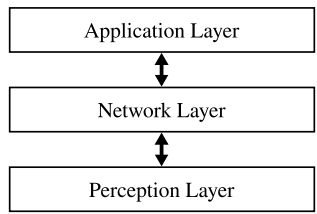

**Fig. 1.** IoT three-layer architecture.

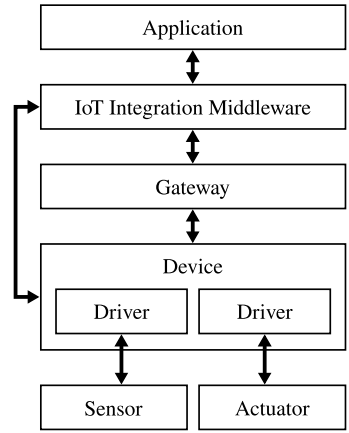

**Fig. 2.** Reference IoT architecture proposed in [19].

*2.1. IoT architecture*

Today there is still no predominant standard or technology used by IoT platforms and devices. Instead, different devices and platforms consider diverse standards and technologies. Therefore, to deploy an IoT system, in many cases users must study, configure, and integrate different architectures and technologies, which can lead to a difficult and time-consuming process.

To solve this problem there has been several efforts lead by information and communications technology (ICT) standards development organizations to create technical specifications for IoT systems. In particular, the most relevant are the IEEE Standard for an Architectural Framework for the Internet of Things [14], the ITU-T Y.2060 Recommendation [15], and the oneM2M standards [16]. In industrial scenarios, for the so-called industrial IoT (IIoT), the Open Platform Communications Unified Architecture (OPC UA) standard [17] and the Industry IoT consortium (formerly the Industrial Internet Consortium) standards [18] are among the most relevant.

In the research literature there has been also many proposed architectures over the past years. One of the simplest is the three-layer model reproduced in Fig. 1, which consists of application, network, and perception layers [2,9]. Several other more complex multi-layered architectures have been also proposed [20,21]. Recently, an abstract reference architecture was introduced in [3] and shown to map well onto different state-of-the-art IoT platform architectures.

The reference architecture proposed in [3], and reproduced in Fig. 2, can be extended considering the models defined in the IEEE Standard for an Architectural Framework for the Internet of Things [14], and in the ITU-T Y.2060 Recommendation [15]. In this way, we can define a more complete reference architecture that is capable of abstracting well the architecture of most of the state-of-the-art IoT platforms. The result is shown in Fig. 3, where security and management blocks have been added, communications networks have been made explicit, and the fundamental blocks that constitute the IoT platform (the IoT integration middleware in Fig. 2) are shown. Such an architecture contains seven different parts: sensors and actuators, IoT platform, communications network, IoT gateway, applications, management, and security. These parts are described below.

*2.1.1. Sensors and actuators*

Sensors are hardware devices that measure physical parameters from the environment (e.g., temperature or humidity) or from other systems (e.g., a current meter). Actuators are hardware devices that carry out actions based on IoT requests (e.g., any device that can be turned on or off). Usually, these are small devices with limited resources, and in the simplest cases the needed processing capabilities may be achieved using a small microcontroller without an operating system [22,23]. When more complex capabilities are needed small operating systems designed specifically for these kind of systems, such as Contiki, TinyOS, and FreeRTOS, are used [24–26]. There exist a vast number of different types of sensors and actuators, these may have different sensed physical parameters, communication and human interfaces, and in some cases they may be also "smart". For a more in-depth review of the types of sensors we refer the reader to the previous works in [27–30].

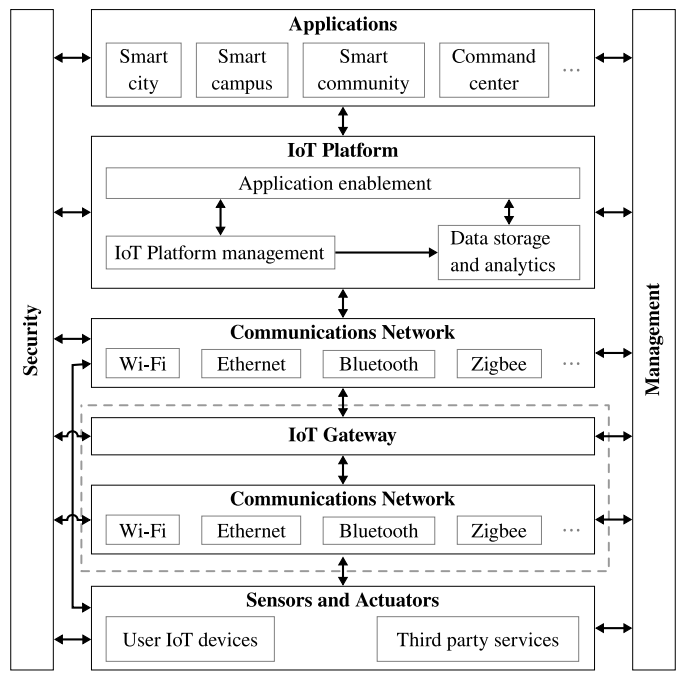

**Fig. 3.** IoT architecture.

For the components shown in the architecture presented in Fig. 3 we distinguish between: *(a)* user IoT devices, which are physical devices deployed by the user, e.g., temperature sensors; and *(b)* third-party services, usually available through a representational state transfer (REST) APIs that provides data on request, e.g., meteorological data provided by government agencies. Sensors can also perform computations to preprocess the captured data prior to transmission. In this case, this is the so-called edge-computing.

### 2.1.2. Communications network

To enable data exchange among sensors, actuators, and the IoT platform and gateway components one or more communications networks must be in place. The communications network must provide the physical, link, and network layers according to the Open Systems Interconnection (OSI) model [31]. The network layer is responsible for providing packet forwarding and routing. In this layer some of the most common protocols are the following:

- *IPv4*: the Internet protocol (IP) version 4 (IPv4) is currently the most used of the IP versions, which are used to address the Internet. It was first implemented in 1983 and uses 32 bit addresses, thus having a total of $2^{32}$ unique addresses, of which $2^{24}$ are reserved for used in private local area networks (LANs). However, due to its low number of addresses, available IPv4 addresses for the Internet are almost depleted. This is the so-called IPv4 address exhaustion problem.
- *IPv6*: To solve the address exhaustion problem, the IP version 6 (IPv6) was developed. IPv6 uses 128 bit addresses, thus being the number of addresses much larger than the one of IPv4.
- *6LoWPAN*: The IPv6 over Low-Power Wireless Personal Area Networks was developed by the Internet Engineering Task Force (IETF) to make an adaptation layer that allows IPv6 with the low-rate wireless personal area networks defined by the IEEE 802.15.4 standard [2].

The physical and link layers are usually defined jointly by communication standards. Usually, IoT devices employ wireless communications, of which there are many different standards currently in use. Some of the most relevant today are presented and compared in Table 1, whose columns show respectively the spatial scope of the network, the name of the technology, its range, the network topology, and the maximum data rate. These technologies and some others were already analyzed in several previous IoT related review papers [2,5,6,11,32]. Furthermore, there are some other specific papers devoted to compare some of these technologies [33–38].

**Table 1**
Overview of some of the most relevant communications technologies for IoT.

| Scope | Technology | Range | Network topology | Max. data rate |
|---|---|---|---|---|
| Proximity | NFC | 10 cm | Point-to-point | 424 kbit/s |
| Personal area network | Bluetooth | ≤100 m | Star | 3 Mbit/s |
|  | Bluetooth Low Energy (BLE) | ≤100 m | Star | 2 Mbit/s |
|  | Thread | 30 m | Mesh | 250 kbit/s |
|  | ZigBee | 30 m (indoors), 100 m (outdoors) | Tree, star, or mesh | 250 kbit/s |
|  | Z-Wave | 40 m (indoors), 300 m (outdoors) | Mesh | 40 kbit/s |
| Local area network | Wi-Fi | ≤70 m (indoors), ≤250 m (outdoors) | Star | 1.2 Gbit/s (Wi-Fi 6) |
| Wide area network | NB-IoT | 10 km | Star | 220 kbit/s |
|  | LoraWan | 20 km | Star of stars | 50 kbit/s (UL), 290 kbit/s (DL) |
|  | LTE-M | 5 km | Star | 1 Mbit/s |
|  | Sigfox | 40 km | Star | 100 bit/s (UL), 600 bit/s (DL) |

#### 2.1.3. IoT gateway

Different IoT devices may use different communication networks, data layer protocols, data formats and, in many cases, the IoT platform may lack direct support for some of these. To solve this problem, a middleware component called "gateway" or "edge gateway" may be introduced between the IoT platform and the IoT devices as shown in Fig. 3. The gateway will provide a common communication network, data layer protocol, and data format with the IoT platform. In real scenarios several gateway devices may be deployed, with each gateway connecting to several IoT devices.

#### 2.1.4. IoT platform

The IoT platform is the central hub of the IoT system, enabling the connected sensors and actuators to be controlled and monitored with the help of applications. As show in Fig. 3, we consider three fundamental building blocks: device management and connectivity, data storage and analytics, and application enablement. These are described below.

- *IoT platform management:* This block provides several management functionalities related to the IoT platform itself. Among others, the functionalities to be provided are the following:
  - *Creation and configuration of user accounts*: enables the creation of accounts for the different users that must interact with the platform, and for each one to configure their authentication and access permissions (see also Section 2.1.6).
  - *Integration and management of devices*: allows the connection of devices to the platform and configures the data to be retrieved (in the case of sensors) or the possible actions to be performed (in the case of actuators). It should be noted that, as explained above for the "communications network" component, different devices or gateways may be connected with different technologies to the IoT platform. Depending on this technology, the means of connecting the devices may be different. Furthermore, if a direct connection to sensors and actuators (i.e., without a gateway) is considered, the underlying data layer protocol and data format may be different. Hence, in this case, the IoT platform must provide a unified interface for accessing sensor data and properties, and for interacting with actuators.
  - *Data and system management*: allows for checking the status of the IoT platform, viewing system logs, creating backup copies of configuration and/or sensor data, and updating software, among others.
  - *Interface to the management and security layers of the IoT system*: as shown in Fig. 3, the security and management blocks, described in Sections 2.1.6 and 2.1.7, respectively, can interact with all other blocks in the system. Thus, the IoT platform management block may provide the means to easily access or interact with those blocks.
- *Data storage and analytics:* This block provides functionalities to the IoT platform to store and access the received sensor data. In addition, it may also provide functionalities to analyze the data. Typically, the platform does not store all the collected data indefinitely, but it is necessary to define the lifetime of the data.
- *Application enablement:* This block provides functionalities for developing and deploying applications for the IoT system. Such functionalities may be an API, integrations with third-party systems, support for rule-based applications, graphical user interface (GUI) development tools for data visualization and control, etc.

In Section 3 we will discuss the available commercial and open-source IoT platforms, their characteristics, and the relevant related works in literature.

#### 2.1.5. Applications

An application is a software component that performs a specific task reading data captured by the sensors and/or controlling the actuators. Examples of applications are:

- A GUI graph that displays the data from a sensor over time.
- A GUI button that allows the user to turn on or off some device.
- A rule that automatically turns on or off a device based on the values provided by another sensor.

One of the most typical and simple applications that are easily supported by many IoT platforms is the so-called rules. These are tasks (e.g., controlling an actuator, performing some operation, sending a warning to the user, etc.) that can be triggered by events such as a sensor returning a value exceeding a certain threshold.

*2.1.6. Security*

The security block is responsible for ensuring confidentiality, integrity and availability. User authentication, user privacy, access control, monitoring and auditing of security events, and management of keys and certificates, are among its functions.

*2.1.7. Management*

The management block allows for management tasks of the IoT system that depend on other blocks than the IoT platform itself (see Fig. 3). Such tasks include fault management, management of the system performance, network management, and remote configuration and update of devices, among others

*2.2. Data layer protocols*

When connecting sensors and actuators to an IoT platform, the communications network blocks shown in Fig. 3 shall provide the required communications capabilities so that these devices may exchange information with the platform. According to the OSI model [31], the corresponding physical, data link, network, and transport protocols must be provided and, on top of them, a data protocol is also required so that IoT devices can exchange data with the IoT platform. In this case, the data protocol implementation is a responsibility of the IoT platform and the IoT devices.

Depending on the considered protocol for data exchange with the IoT platform, three basic models can be followed [11], which are respectively illustrated in Figs. 4(a) and 4(b), 4(c), and 4(d):

1. The *publish model* is a client–server model where a client device sends directly some data to a server. An example of this model is shown in Fig. 4(a), where a sensor sends some data to the IoT platform with a "publish" message. In this case, the message carries the identifier $i$ (e.g., a label or a uniform resource identifier (URI)) to identify the resource in the IoT platform, and some associated data $d$. The same model could be used by the IoT platform to control actuators as shown in the example in Fig. 4(b).
2. The *publisher-subscriber model* involves three types of entities (see Fig. 4(c)): "publishers", "subscribers", and a server, which is usually referred to as "broker". In this model, "publishers" send publish messages to the broker containing data $d$ together with an identifier $i$ (e.g., a name or topic). A "subscriber" wishing to receive data for that identifier must first send a corresponding subscribe request to the "broker". The "broker" is a middleware in charge of receiving messages from "publishers" and sending them to the appropriate "subscribers". As shown in Fig. 4(c), this model is typically used for sending data from the sensors to the IoT platform, where the IoT platform acts as a "subscriber" and the sensors act as "publishers", although it may be also used for the case of the IoT platform to send data to actuators, in which case the actuators would act as "subscribers" and the IoT platform as the "publisher".
3. The *polling model* is a client–server model where the client device requests data from a server. An example of this model is shown in Fig. 4(d), where the IoT platform requests data for a given identifier $i$, and the sensor responds with the corresponding data $d$.

The most prominent software architectural styles that use these models are:

- representational state transfer (REST) is a software architecture which usually uses the Hypertext Transfer Protocol (HTTP)'s GET, POST, PUT and DELETE methods to recover, create, update, or delete resources. In an IoT system, sensors, actuators, and the IoT platform may have REST APIs to recover data or send commands. In this case, the communication models, among the ones presented above, will be the publish and the polling ones.
- event driven architecture (EDA) is a software architecture based on events, in which an event may be described as any change in the system. Parts of the system may "listen" to certain events and react accordingly when the events are triggered [39]. In this case the communication model followed is the publisher-subscriber one.

Some of the most relevant application layer protocols for IoT are the following:

1. *Matter* [40] is a recent protocol being developed by a working group within the Connectivity Standards Alliance (CSA, formerly the Zigbee Alliance). This working group is promoted by important companies such as Amazon, Apple, Google, and Comcast, among many others. The objective of the working group is the development of a new, royalty-free connectivity standard to increase compatibility among smart home products, with security as a fundamental design principle, simplifying development for manufacturers and increasing compatibility for consumers [41].

An Overview of IoT Architectures, and Existing Open-Source Projects

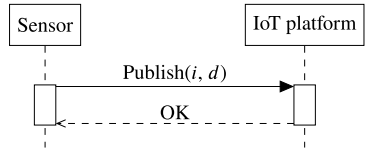

(a) Publish model: from a sensor to the IoT platform.

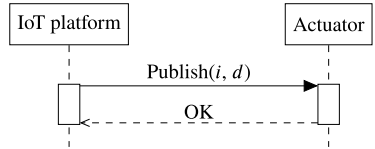

(b) Publish model: from the IoT platform to an actuator.

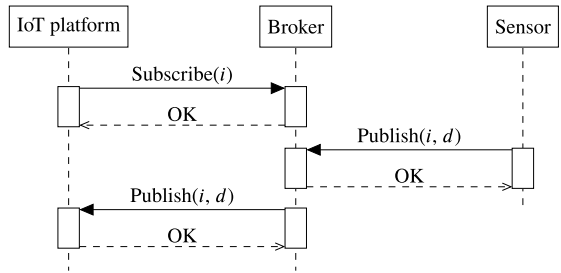

(c) Publisher-subscriber model: information exchange between a sensor and the IoT platform.

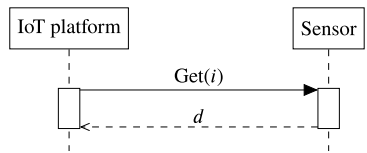

(d) Polling moddel: from the IoT platform to a sensor.

**Fig. 4.** Illustration of data exchange models.

2. The *Hypertext Transfer Protocol (HTTP)* [42] is the base of the communications for the World Wide Web. It is a generic stateless protocol which follows the client–server model, and can be used for many different tasks beyond its use in the World Wide Web. In the IoT world, we can use HTTP to exchange data with a plain publish model or with a polling model.
3. The *Constrained Application Protocol (CoAP)* [43] is a specialized protocol intended for use in constrained environments (e.g., low-power devices or lossy networks). According to its specification, CoAP is designed for machine to machine (M2M) applications such as smart energy and building automation [43]. CoAP follows a client–server model, similar to HTTP, and in fact, it can be easily adapted to communicate with HTTP endpoints. Hence, in the same way as with HTTP, CoAP can be used to exchange data with a plain publish model or with a polling model.
4. *Message Queuing Telemetry Transport (MQTT)* [44] is a lightweight publisher-subscriber protocol. According to its specification, MQTT is ideal for M2M and IoT contexts where a small code footprint is required and/or network bandwidth is at a premium [44].
5. *Advanced Message Queuing Protocol (AMQP)*. There are two widely used versions of AMQP: AMQP 0-9-1 [45], and AMQP 1.0 [46]. For AMQP it is important to consider its version because their specifications differ significantly. On the one hand, the AMQP 0-9-1 protocol follows a publish–subscriber model, defining the interoperability clients and messaging middleware servers ("brokers"), and thus the only supported data exchange model is the publish–subscriber one. Compared to MQTT, an AMQP 0-9-1 broker uses queues to store received messages from which the consumers should read them (usually one consumer per queue), but MQTT does not use queues, instead it just simply forwards the received messages to the corresponding subscribers. On the other hand, the AMQP 1.0 protocol simply specifies the reliable exchange of business messages between two parties, without requiring a broker, although one may be used. Thus, the AMQP 1.0 protocol can be used to exchange data with a plain publish model, a publisher-subscriber model, or a polling model.

It should be noted that, when transmitting data with these protocols, another security protocol, such as Transport Layer Security (TLS) or Datagram TLS (DTLS), must be used on top in order to provide privacy, integrity, and authenticity of the communications. A more in-depth review of data and security protocols for IoT can be found in several other previous review papers [2,5,6,11,47].

*2.3. Data formats*

Data layer protocols allow for exchanging information between the IoT platform and IoT devices. However, these protocols do not impose any restriction on the data type (i.e., text or binary) and format (i.e., the structure). Hence, the type and format sent by each IoT device may be different, and, on the same way, a given IoT platform may only accept data from sensors with limited types and formats. Thus, for connecting a given IoT device to an IoT platform we must be sure that the device data is transmitted in a format accepted by the IoT platform, and configure the IoT platform so that the desired data is interpreted correctly.

Usually, data is encoded using one of the so-called data-serialization formats, which allow complex data to be encoded in binary or text representations. Among them, some of the most popular are:

- *JavaScript Object Notation (JSON)* [48] is a simple text format to serialize data as a collection of name–value pairs (an object in the JSON nomenclature), where each value may be a string, a number, `true`, `false`, `null`, another object, or an array.
- *Concise Binary Object Representation (CBOR)* [49] is a binary data format based on JSON whose design goals are the possibility to be implemented with an extremely small code size, a fairly small message size, and extensibility.
- *comma-separated values (CSV)* is a text format that serializes values separated with commas. The 2005 Request for Comments (RFC) 4180 [50] provides a formal definition of the CSV format.
- *Extensible Markup Language (XML)* [51] is a simple and flexible text format widely used over the Internet.
- *Protocol Buffers* [52] is a data format developed by Google for serializing structured data into a compact binary representation.
- *YAML* [53] is a text-based data serialization language commonly used for configuration files, but also for Internet messaging and object persistence. As of version 1.2, YAML is officially a superset of the JSON format, thus JSON documents are compliant with YAML 1.2.

Previous works in the literature have compared the performance of several of these data-serialization formats (and also others). In particular, the works in [54–57] provide comparisons of different data serialization formats, assessing their serialization speed and data size.

## 3. IoT platforms

Below we describe some of the most relevant IoT platforms available today. For this purpose, we divide them in two categories: commercial platforms, and free and open-source platforms.

*3.1. Commercial IoT platforms*

We distinguish between consumer-oriented platforms and enterprise platforms.

*3.1.1. Consumer-oriented platforms*

These platforms are consumer-centric and provide plug-and-play functionality with many devices on the market, allowing users to communicate with and control them. However, the services, options, and functionalities provided may be limited, as these platforms are aimed at non-enterprise users. The IoT devices available for these platforms are usually household items such as lights, cameras, energy monitoring sockets, or other appliances. Among these platforms, two of the most relevant are Apple HomeKit [58] and SmartThings [59].

*3.1.2. Enterprise platforms*

These platforms are targeted to professional and industrial environments, and thus provide highly scalable and customizable solutions, while also requiring specialized developers to configure and deploy IoT systems with them. These are general-purpose IoT platforms that allow for connecting many devices, and collecting, storing, and analyzing their data for a wide range of domains. Being enterprise platforms, they provide robust security mechanisms, scalability, and advanced data analytics based on modern methods such as machine learning. These platforms usually run on a cloud,[1] and the user must pay a periodical subscription to use them. Some of the most popular platforms are Amazon Web Services (AWS) IoT [60], Google Cloud IoT [61], IBM Cloud [62], Microsoft Azure IoT [63], and ThingWorx [64].

As shown in Section 1, there are already many works in the literature comparing these platforms.

---

[1] Hence, these platforms are provided as SaaS, infrastructure as a service (IaaS), and platform as a service (PaaS).

*3.2. Free and open-source IoT platforms*

A number of free and open-source IoT platforms are available. Analogous to enterprise platforms, they allow IoT devices to be connected and their data to be collected and stored. However, unlike enterprise platforms, free and open-source platforms can be very heterogeneous, and in some cases the data analytics, scalability, or security functionalities may be inferior. Some of the most popular examples, which we consider for comparison, are Eclipse Kapua [65], Home Assistant [66], Mainflux [67], openEMS [68], openHAB [69], OpenRemote [70], SiteWhere [71], and ThingsBoard [72].

In contrast to the case of commercial IoT platforms, there are not many works in the literature that have considered open-source IoT platforms for comparison. From the works presented above, open-source IoT platforms are only considered in [10] (ThingsBoard and SiteWhere), and in [11] (openHAB).

For the purpose of comparing and describing these different IoT platforms, similarly to previous works [7,8,11], we will consider the following technical characteristics: º

- *License:* The type of open-source license used by the platform. The license specifies the terms of use of the source code and software. For a software to be considered open-source the license of the code must comply with the terms defined by the Open Source Initiative [73]. The most common licenses for the IoT software platforms and projects discussed in Sections 3.2 and 4 are:
    - Apache License, v. 2.0 [74].
    - Eclipse Public License, v. 2.0 (EPLv2) [75].
    - GNU Affero General Public License, version 3 (AGPLv3) [76].
    - GNU General Public License, version 2 (GPLv2) [77] and version 3 (GPLv3) [78].
    - MIT License [79].
- *Focus:* The application domain on which the platform is focused, distinguishing three different targets: *(1)* home automation, *(2)* energy management, and *(3)* general-purpose IoT applications.
- *Programming language:* The programming language used for the platform backend (i.e., the core of the platform). For the frontend, i.e., the user interface (UI), in most cases HTML together with Javascript is used, and therefore will not be considered.
- *Supported integrations:* We consider two types of integrations:
    - *Integrations with commercial off-the-shelf IoT devices:* Each of these devices may require a different data layer protocol and use different data formats, which the platform must be aware of in order to integrate them.
    - *Generic integrations:* Integrations using existing protocols such as the ones described in Section 2. In this case, devices using the supported protocols may be integrated with the platform, but depending on the IoT platform some limitations may arise, such as the allowed data format.
- *Supported databases:* The databases supported by the IoT platform to store the data from the IoT devices. Nowadays, there are different types of databases in use:
    - *Relational databases* are based on the relational model [80], where data is stored into one or more tables, and SQL is used to query the data. Some of the most popular open-source relational databases are PostgreSQL [81], MySQL [82], MariaDB [83], and SQLite [84].
    - *NoSQL databases* are not based on the relational model and data is queried using a non-SQL language, usually specific to each database. Some notable open-source NoSQL databases are Apache Cassandra [85], Apache HBase [86], and Apache CouchDB [87]. It is worth noting that the two largely popular databases Elasticsearch [88] and MongoDB [89] were open-source projects up to 2018 and 2021, respectively, when they switched their licenses to the Server Side Public License (SSPL), a source-available but not open-source license (according to the definition of open-source of the Open Source Initiative [73]).
    - *Time-series databases* are designed to efficiently store time-series data, i.e., tuples of a timestamp with other additional values. Hence, they are commonly used in IoT platforms to store sensor data. Some notable open-source time-series databases are InfluxDB [90], TimescaleDB [91], and OpenTSDB [92]. In the literature, [93,94] provide extensive surveys about time-series databases.
- *Automations* define actions that are triggered when an event occurs or a condition is met. Automations may be specified using simple rules based on values provided by sensors. There are also platforms that enable the definition of automations based on flow-based frameworks such as Node-RED [95].
- *Integrated data analytics:* The functionalities provided by the IoT platform to perform any type of statistical analysis of the data received from the IoT sensors.
- *Integrated data visualization:* The capabilities provided by the UI of the IoT platform to display the data received from the IoT sensors.

Tables 2 and 3 show the comparison of the considered open-source IoT platforms based on the characteristics described above. In these tables, the first column contains the characteristics that will be assessed, and the following columns contain the assessment for the IoT platforms considered (one IoT platform per column). For each platform assessed we also show the version and corresponding release date that we have considered for the comparison.

**Table 2**
Free and open-source IoT platforms comparison (part 1 of 2).

| IoT platform | Eclipse Kapua | Home Assistant | Mainflux | openEMS |
|---|---|---|---|---|
| Considered version | 1.6.0 (Jul 2022) | 2022.7 (Jul 2022) | 0.13.0 (Apr 2022) | 2022.7.0 (Jul 2022) |
| License | EPLv2 [75] | Apache 2.0 [74] | Apache 2.0 [74] | AGPLv3 [76] |
| Focus | Generic IoT applications | Home automation | Generic IoT applications | Energy management |
| Programming language | Java | Python | Go | Java |
| Supported integrations | Currently only MQTT using Eclipse Kura (or the Eclipse Kura protocol format) | +1800, also including generic integrations such as command line, HTTP, MQTT, etc. [96] | Generic integrations such as HTTP, MQTT, WebSocket, or CoAP | +60. Generic integrations for M-Bus, ModBus, and 1-wire |
| Supported databases | Elasticsearch | MariaDB, MySQL, PostgreSQL, and SQLite | Cassandra, InfluxDB, MongoDB, and PostgreSQL | InfluxDB |
| Automations | External applications may subscribe to the Eclipse Kapua MQTT broker to listen to the received data and react accordingly | Trigger-action automations can be created via the UI or with the YAML language (there are pre-defined automations named blueprints). Flow-based rules can also be created using Node-RED | An event-based system is used to communicate changes in devices and their data. External applications need to be developed to listen to these events and react accordingly | A publisher-subscriber system is provided to notify the user about updated data. External applications need to be developed to subscribe to this system and react accordingly |
| Integrated data analytics | N/A | Simple descriptive statistics using the "Statistics" integration or by querying the database with the "SQL" integration | N/A | N/A |
| Integrated data visualization | N/A | The UI supports multiple dashboards in which different data visualization and control panels can be placed. | N/A | The UI provides instantaneous and time plots of the power produced and consumed in the monitored electric installation. |

**Table 3**
Free and open-source IoT platforms comparison (part 2 of 2).

| IoT platform | openHAB | openRemote | SiteWhere | ThingsBoard |
|---|---|---|---|---|
| Considered version | 3.3.0 (June 2022) | 3.0 (Jul 2022) | CE 3.0.5 (Jul 2021) | CE 3.4 (Jul 2022) |
| License | EPLv2 [75] | AGPLv3 [76] | Common Public Attribution License (CPAL), v. 1.0 [97] | Apache 2.0 [74] |
| Focus | Home automation | Generic IoT applications | Generic IoT applications | Generic IoT applications |
| Programming language | Java | Java | Java | Java |
| Supported integrations | +2000. The only generic integration is MQTT [98]. | Generic integrations such as HTTP, TCP, MQTT, etc. [99] | Generic integrations such as HTTP, TCP, MQTT, etc. [100] | Generic integrations: HTTP, MQTT, CoAP, LwM2M, and SNMP [101]. |
| Supported databases | Amazon DynamoDB, InfluxDB, MongoDB, or any other through JDBC or the Java Persistence API | PostgreSQL | InfluxDB and Warp 10 | Cassandra, PostgreSQL, and TimescaleDB |
| Automations | Trigger-action automations can be created via the user interface, or with the Xbase language. Flow-based rules can also be created using Node-RED | Trigger-action automations and flow-based rules can be created via the user interface. | Automations can be created using the WSO2 Siddhi Query Language | Flow-based rules can be created via the user interface. |
| Integrated data analytics | Statistics can be obtained querying the database with the "DBQuery" integration | N/A | N/A | Analysis of incoming telemetry data can be performed using Apache Kafka [102] |
| Integrated data visualization | The UI shows sensors and actuators grouped by location and type. For each sensor, an historic data plot is provided | For each sensor, the attributes, location, and historic data is shown. An "insights" tab is also available to plot historic data of several sensors | N/A | The UI allows the creation of multiple dashboards in which different data visualization and control panels can be placed. |

**Table 4**

Summary of IoT open-source platforms and projects. For each project the table details the type, the name, the developer, the first release date, the license (or licenses if there are different parts with different licenses), the commercial support, the number of GitHub stars in its official GitHub repository (or N/A if no official GitHub repository exists), and the number of search results in GitHub when searching for repositories using the project name. GitHub data was collected in August 2022. For each type, projects are sorted according to their number of stars in GitHub.

| Type | Software project | Developer | First release date | License | Commercial support | GitHub stars | GitHub search results |
|---|---|---|---|---|---|---|---|
| DB (Relational) | PostgreSQL [81] | PostgreSQL Global Development Group | 8 Jul. 1996 | PostgreSQL [103] | External | 10 700 | 82 753 |
| | MySQL [82] | Oracle | 23 May 1995 | GPLv2 [77] | ✓ | 8 086 | 305 629 |
| | MariaDB [83] | MariaDB Foundation | 29 Oct. 2009 | GPLv2 [77] | ✓ | 4 383 | 11 718 |
| | SQLite [84] | D. Richard Hipp | 17 Aug. 2000 | Public domain | ✓ | 2 675 | 75 720 |
| DB (noSQL) | Apache Cassandra [85] | Apache Software Foundation | Jul. 2008 | Apache 2.0 [74] | ✗ | 7 442 | 15 423 |
| | Apache CouchDB [87] | Apache Software Foundation | 2005 | Apache 2.0 [74] | ✗ | 5 392 | 6 076 |
| | Apache HBase [86] | Apache Software Foundation | 28 Mar. 2008 | Apache 2.0 [74] | ✗ | 4 596 | 7 570 |
| DB (time-series) | InfluxDB [90] | InfluxData | 24 Sep. 2013 | MIT [79] | ✓ | 23 910 | 8 598 |
| | TimescaleDB [91] | Timescale, Inc. | 1 Nov. 2018 | Apache 2.0 [74] | ✓ | 13 426 | 440 |
| | OpenTSDB [92] | B. Sigoure et al. | 2010 | GPLv3 [78] | ✗ | 4 708 | 4 708 |
| OPC UA | open62541 [104] | J. Pfrommer, S. Profanter, et al. | 11 May 2015 | MPL v2.0 [105] | ✗ | 1 848 | 153 |
| | FreeOpcUa [106] | R. Alexander, O. Roulet-Dubonnet, M. Bec, et al. | 2013 | GPLv3/LGPLv3 [78,107] | ✗ | 576 | 27 |
| | S2OPC [108] | Systerel | 17 Feb. 2017 | Apache 2.0 [74] | ✓ | 10 | 6 |
| oneM2M | OpenMTC [109] | R. Steinke, C. Klopp, et al. | 4 Mar. 2018 | EPLv1 [110] | ✗ | 42 | 14 |
| | Eclipse OM2M [111] | Eclipse Foundation | 8 Apr. 2015 | EPLv2 [75] | ✗ | N/A | 7 |
| Gateway | EdgeX Foundry [112] | The Linux Foundation | 2 Oct. 2017 | Apache 2.0 [74] | External | 1 060 | 48 |
| | Eclipse Kura [113] | Eclipse Foundation | 7 Apr. 2014 | EPLv2 [75] | ✗ | 424 | 64 |
| Integration tool | Apache Kafka [114] | Apache Software Foundation | Jan. 2011 | Apache 2.0 [74] | ✗ | 22 657 | 75 785 |
| | Node-RED [95] | JS Foundation | 16 Oct. 2013 | EPLv2 [75] | ✗ | 15 042 | 16 231 |
| | Telegraf [115] | InfluxData, Inc. | 19 Jun. 2015 | MIT License [79] | ✓ | 11 789 | 3 018 |
| | Fluentd [116] | Treasure Data, Inc. et al. | 10 Oct. 2011 | Apache 2.0 [74] | External | 11 377 | 3 857 |
| | Fluent Bit [117] | Treasure Data, Inc. et al. | 30 Dec. 2016 | Apache 2.0 [74] | External | 3 788 | 492 |
| | collectd [118] | F. Forster, et al. | 8 Jul. 2005 | MIT/GPLv2 [77,79] | ✗ | 2 769 | 1 847 |
| | TCollector [119] | B. Sigoure et al. | 2010 | Apache 2.0 [74] | ✗ | 498 | 28 |
| | Eclipse Hono [120] | Eclipse Foundation | 12 Feb. 2018 | EPLv2 [75] | ✗ | 378 | 30 |
| | WSO2 SI [121] | WSO2 LLC | 17 Feb. 2020 | Apache 2.0 [74] | ✓ | 70 | 7 |
| IoT platform | Home Assistant [66] | P. Schoutsen et al. | 17 Sep. 2013 | Apache 2.0 [74] | ✗ | 54 247 | 13 808 |
| | ThingsBoard [72] | ThingsBoard, Inc. | 6 Dec. 2016 | Apache 2.0 [74] | ✓ | 12 106 | 818 |
| | Mainflux [67] | Mainflux Labs | 15 Mar. 2018 | Apache 2.0 [74] | ✓ | 1 836 | 103 |
| | openHAB [69] | openHAB Foundation et al. | 2010 | EPLv2 [75] | ✗ | 1 172 | 2 732 |
| | SiteWhere [71] | SiteWhere LLC | 2014 | CPAL 1.0 [97] | ✓ | 901 | 88 |
| | openRemote [70] | OpenRemote, Inc. | 2016 | AGPLv3 [76] | ✓ | 615 | 63 |
| | openEMS [68] | OpenEMS Association e.V | 21 Jun. 2017 | AGPLv3 [76] | ✗ | 307 | 73 |
| | Eclipse Kapua [65] | Eclipse Foundation | 1 Dec. 2017 | EPLv2 [75] | ✗ | 192 | 6 |
| Visualization | Grafana [122] | Grafana Labs | 19 Jan. 2014 | AGPLv3 [76] | ✓ | 50 246 | 16 413 |
| | Metabase [123] | Metabase | Jun. 2015 | AGPLv3 [76] | ✓ | 29 316 | 1 166 |
| | Redash [124] | Redash, Ltd. | 25 Feb. 2014 | BSD 2-clause [125] | ✓ | 21 506 | 618 |

## 4. Free and open-source IoT-related projects

Apart from the IoT platforms described above, there exist also many other software projects that, although they are not IoT platforms, they provide important functionalities to be considered as building blocks for developing a custom IoT platform. Some of them are detailed below.

*4.1. OPC UA*

As said in Section 2.1, The Open Platform Communications Unified Architecture (OPC UA) is an standard developed by the OPC Foundation for the industrial IoT [17]. The OPC UA standard defines two types of entities, clients and servers, which can communicate following a publish model using request and response messages, and also following a publisher-subscriber model by a client subscribing to an OPC UA server. There exist many open-source implementations of both OPC UA clients and servers, for example open62541 [104], S2OPC [108], and FreeOpcUa [106], among others.

*4.2. oneM2M*

In section Section 2.1 the oneM2M standards were introduced, these are being developed with the aim to create a set of common technical specifications to achieve interoperable and secure IoT systems [16]. There exist several open-source projects which provide implementations of the oneM2M standards, for example, Eclipse OM2M [111] and OpenMTC [109], among others.

*4.3. FIWARE*

FIWARE [126,127] is an IoT framework supported by the European Commission. Its main objective is to provide a set of common specifications and open-source platform components to accelerate the development of IoT solutions. FIWARE is specifically used in the domain of smart cities, being already deployed in more than 40 cities around the world [128], whereas other domains include smart energy, smart agriculture, and smart-industry [126].

*4.4. Gateways*

As explained in Section 2.1.3, in some cases a gateway may be introduced between the IoT platform and the IoT devices. EdgeX Foundry [112], developed under the Linux Foundation, and Kura [113], developed under the Eclipse Foundation, are two popular open-source IoT edge frameworks that enable sensor data to be captured and processed on the edge gateway and then sent to an IoT platform or other service.

*4.5. Data collection and integration tools*

There exist many tools that allow for collecting, processing, and moving data and metrics from different sources and protocols. These tools enable one of the most important features of IoT platforms or IoT systems, which is the integration with different data sources. Thus, these tools should be considered when designing an IoT system. The most relevant data collection and integration tools are Apache Kafka [114], collectd [118], Eclipse Hono [120], Fluent Bit [117], Fluentd [116], Node-RED [95], TCollector [119], Telegraf [115], and WSO2 Streaming Integrator (WSO2 SI) [121].

*4.6. Data visualization*

There are also tools for data visualization which support writing queries to a database and plotting the resulting data in many different ways. Therefore, these tools should be considered when there is a need of visualizing the IoT sensors data in complex ways. Popular data visualization tools are Grafana [122], Metabase [123], and Redash [124].

**5. Summary of IoT open-source platforms and projects**

In Table 4 we provide a detailed list of the open-source software projects discussed in previous sections. Table 4 contains the seven columns described below:

1. The type of project. We consider data-bases (DBs) and IoT platforms as discussed in Section 3.2, and the different categories of IoT projects discussed in Section 4. Note that each row details a single project, but a single type may contain several projects, being the different types separated by horizontal rules.
2. The current developer of the project.
3. The date of the first public release of the project.
4. The license used by the project. In some cases a project may have components with different licenses, in that case we show the different licenses separated with a slash. Note that in many cases the developers may also offer proprietary versions of the project with additional features not present in the open-source version.
5. Commercial support readily available for the project. We distinguish the following cases:
   (a) We use the "✓" mark to indicate that there is direct commercial support provided by the developers of the project. In this case, the developers may also provide some proprietary version of the software with additional features.
   (b) If there is no support by the developers of the project, but the project indicates that there are other external companies providing commercial support, we indicate it with the word "external".
   (c) In other case, we indicate with the "✗" mark that there is no commercial support. Note that in this case there may still be commercial support provided by external companies, but this is not acknowledged by the developers.
6. The number of stars the project has on GitHub (https://github.com/). GitHub is today one of the most popular repositories for free and open source software. Most of the analyzed projects are officially on GitHub, i.e., the project code has been uploaded to GitHub by the developers. GitHub code repositories can be "starred" by users, and thus the number of "stars" of a repository is a good metric to indicate its popularity.

7. The number of repositories in GitHub when searching using the project name as the search string. This is another metric that provides insight of the popularity of each software. Note that for the Apache projects we did not include the term "Apache" since this is usually omitted in references from other projects and there is no ambiguity with other projects.

## 6. Conclusion

In this article, we have first provided a comprehensive overview of key IoT elements, consisting of the IoT architecture, the data layer protocols, and data formats. Regarding the IoT architecture, based on previous literature and standards, we have proposed a detailed reference IoT architecture and have described its parts. Next, for data layer protocols we have explained some of the most popular data layer protocols used for IoT systems. Here, we noted an important point, that data layer protocols do not impose restrictions on the data format, so the data exchanged by IoT sensors and IoT platforms must be in some format accepted by both. Hence, we have reviewed some of the most popular formats for data exchange in IoT systems. Finally, in real-world applications, IoT devices may use different communications networks, data layer protocols, and data formats, which may not be compatible with some IoT platforms, for this case the IoT architecture may include a gateway component which provide a common communication network, data layer protocol, and data format with the IoT platform. Thus, this overview have provided some of the most important foundational knowledge for deploying any IoT system.

Following the overview of the key elements of IoT, we have presented a survey of IoT platforms. For this purpose they have been divided into two categories: commercial platforms and free and open source platforms. In the case of the former there are already many previous works in the literature that provide a comparison of these platforms, so we have only included a basic description and cited the relevant works. However, there are not many previous works in the literature about free and open source platforms. Therefore, in this paper we have provided an overview of free and open source IoT platforms, detailing some of the most popular ones today. It has been shown that, unlike enterprise platforms, free and open source IoT platforms are very heterogeneous, with large differences between them in terms of functionalities. Therefore, having an overview is a useful reference when deciding which platform to use. There are also many other free and open source software projects that, although they are not IoT platforms, can be used as building blocks for IoT systems, so we have provided an overview of some of the most relevant ones. Finally, we have compiled a table summarizing all the free and open source software IoT platforms in addition to other related projects.

**Declaration of competing interest**

The authors declare that they have no known competing financial interests or personal relationships that could have appeared to influence the work reported in this paper.

**Data availability**

No data was used for the research described in the article.

**Acknowledgments**

This research/work has been supported by GAIN (Galician Innovation Agency) and the Regional Ministry of Economy, Employment and Industry, Xunta de Galicia under grant COV20/00604 through the ERDF Galicia 2014-2020; and by grant PID2019-104958RB-C42 (ADELE) funded by MCIN/AEI/10.13039/501100011033. Funding for open access charge: Universidade da Coruña/CISUG.